\documentclass[letter]{aa}
\usepackage{graphicx}
\usepackage{natbib}

\begin{document}
\title{Rotation at 1122 Hz and the neutron star structure}

\author{M. Bejger\inst{1,2}
\and P. Haensel\inst{1} \and J.L. Zdunik\inst{1}
}
\institute{
N. Copernicus Astronomical Center, Polish
           Academy of Sciences, Bartycka 18, PL-00-716 Warszawa,
           Poland
           {\em jlz@camk.edu.pl}}
\institute{
N. Copernicus Astronomical Center, Polish
           Academy of Sciences, Bartycka 18, PL-00-716 Warszawa,
           Poland
           \and
           LUTh, UMR 8102 du CNRS, Pl. Jules Janssen, 92195
            Meudon, France\\
           {\tt bejger@camk.edu.pl, haensel@camk.edu.pl,
           jlz@camk.edu.pl}}

\offprints{J.L. Zdunik}
\date{Received 11 December 2006 / Accepted 26 January 2007}
\abstract{}{Recent observations of XTE J1739-285 suggest that
it contains a neutron star rotating at 1122 Hz. Such rotation imposes
bounds  on the structure of neutron star in
XTE J1739-285. These bounds may be  used to constrain
poorly known equation of state of dense matter at densities
$\ga 10^{15}~{\rm g~cm^{-3}}$.}{One-parameter families of
stationary configurations rotating rigidly at 1122 Hz are
constructed, using a precise 2-D code solving
Einstein equations.  Hydrostatic equilibrium solutions
are tested for stability with respect to axi-symmetric perturbations.
A set of ten diverse EOSs of neutron stars is considered. Hypothetical
strange stars are also studied.}
{For each EOS, the family of possible
neutron star models is limited by the mass shedding limit,
corresponding to maximum allowed equatorial radius, $R_{\rm
max}$, and by the instability with respect to the
axi-symmetric perturbations, reached at the minimum allowed
equatorial radius, $R_{\rm min}$. We get  $R_{\rm min}\simeq
10-13$~km, and $R_{\rm max}\simeq 16-18$~km, with allowed mass
$1.4-2.3~M_\odot$. Allowed stars with hyperonic or exotic-phase core are
supramassive and have  a very narrow mass range. Quark star with accreted
crust might be allowed, provided such a model is  able to reproduce
 X-ray bursts from  XTE J1739-285.}{}

\keywords{dense matter -- equation of state --
stars: neutron -- stars: rotation}

\titlerunning{Neutron star rotating at 1122Hz}
\maketitle
%
\section{Introduction}
\label{sect:introd}
Because of their  strong gravity,
neutron stars can be very rapid rotators; theoretical studies
indicate that they could rotate at submillisecond periods,
i.e., at  frequency $f=1/{\rm period}>$1000 Hz (e.g., \citealt{CST1994},
\citealt{Salgado1994}). During 24  years after its detection,
the first millisecond pulsar B1937+21,
rotating at $f=641$ Hz  \citep{Backer1982}, remained the most rapid one.
In 2006, a more rapid pulsar J1748-2446ad, rotating at $f=716$ Hz,
 was detected \citep{Hessels2006}. However, such frequencies are too low to
significantly affect structure of neutron stars with $M>1M_\odot$
\citep{STW1983}. They belong to a {\it slow rotation} regime,
because $f$ is significantly smaller than the mass shedding
(Keplerian) frequency $f_{\rm K}$.  Under the slow rotation
regime, effects of rotation on neutron star structure are
$\propto (f/f_{\rm K})^2\ll 1$. To enter the rapid rotation
regime for $M>1M_\odot$, one needs submillisecond pulsars with
$f>1000$ Hz.

Very recently \citet{Kaaret2006} reported a
discovery of oscillation frequency $f=1122$ Hz in an X-ray burst
from the X-ray transient,  XTE J1739-285. According to
\citet{Kaaret2006} "this oscillation frequency suggests that
XTE J1739-285 contains the fastest rotating neutron star yet
found". The very fact that stable rotation of neutron star
at $f>1000$ Hz exists means that instabilities that could
set in at lower $f$ (r-mode instability, other gravitational-radiation-reaction
instabilities) are effectively damped.   In the present Letter
 we derive, using  precise 2-D
calculations of neutron stars with different
EOSs, constraints on neutron star in  XTE J1739-285, which
result from the condition of stable rotation at $f=1122$ Hz.

Neutron star models at $f=1122$ Hz are
analyzed in Sect.~\ref{sect:NS}, where we also derive
constraints on neutron star parameters resulting from the
condition of stable rotation. Possibility of strange (quark) star
in XTE J1739-285 is briefly discussed in Sect.\ \ref{sect:QS}.
Sect.\ \ref{sect:discuss} contains discussion of our results
and conclusions.
\section{Neutron stars at 1122 Hz}
\label{sect:NS}
\begin{figure}[h]
\centering
\resizebox{3.5in}{!}{\includegraphics[]{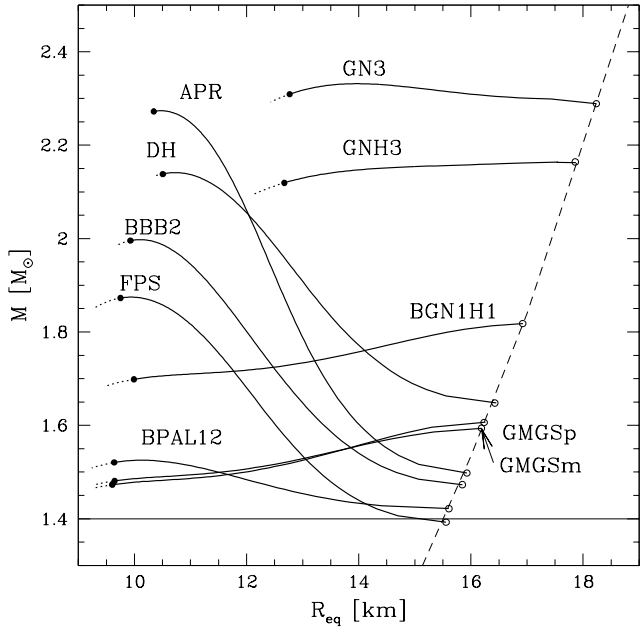}}
\caption{Gravitational mass, $M$, vs. circumferential equatorial radius,
$R_\mathrm{eq}$,    for neutron stars stably rotating at $f=1122$
Hz, for ten EOSs labeled  as in Tab.\ \ref{tab:RminRmax}.
Small-radius termination by filled circle: setting-in of
instability with respect to the axi-symmetric perturbations.
Dotted segments to the left of the filled circles: configurations unstable with
respect to those perturbations.
 Large-radius termination by an open circle: the mass-shedding
instability. The mass-shedding points are very well fitted by the
dashed curve  $R_{\rm min}=15.52\;(M/1.4 M_\odot)^{1/3}\;$km.
For further explanation see the text.
 }
\label{fig:MR-NS}
\end{figure}

\begin{figure}[h]
\centering
\resizebox{3.5in}{!}{\includegraphics[]{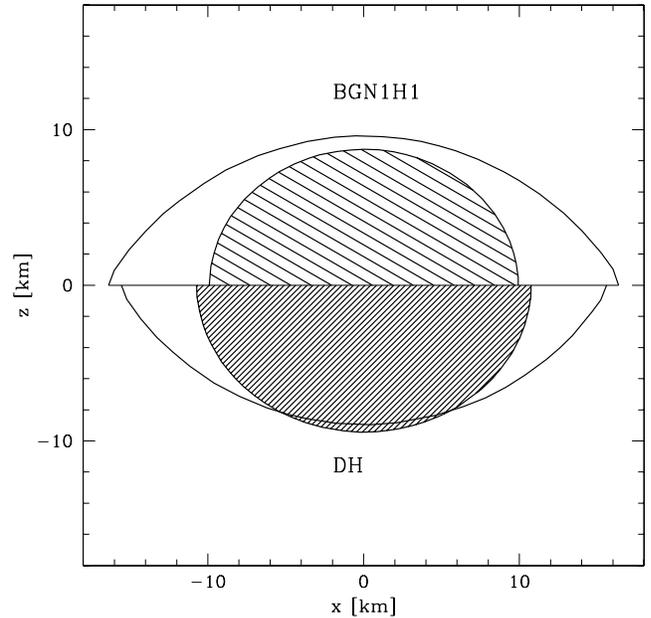}}
\caption{Cross section in the plane passing through the
rotation axis of neutron stars rotating at 1122 Hz,
for the BGN1H1 EOS ( $z>0$) and
DH EOS ($z<0$). Blank area with
solid line contour - $R_{\rm eq}=R_{\rm max}$; hatched area -
$R_{\rm eq}=R_{\rm min}$. The coordinates $x$ and $z$ are
defined as $x=r\mathrm{sin}\theta \mathrm{cos}\phi$,
$z=r\mathrm{cos}\theta$, where $r$ is radial coordinate in the space-time
metric.
 }
\label{fig:shapesNS}
\end{figure}
The  stationary configurations of rigidly rotating neutron
stars have been computed in the framework of general relativity
by solving the Einstein equations for stationary axi-symmetric spacetime
(see \citealt{BGSM1993} or \citealt{GourgSS1999}). The
numerical computations have been performed  using  the {\tt
Lorene/Codes/Rot\_star/rotstar} code from the LORENE library
({\tt http://www.lorene.obspm.fr}). One-parameter
families of stationary 2-D configurations were
calculated for ten EOSs of neutron-star matter, listed in
Table\ \ref{tab:RminRmax}. Stability with
respect to the mass-shedding from the
equator implies that at a given gravitational mass $M$
the circumferential equatorial radius $R_{\rm eq}$
should be smaller than $R_{\rm max}$ which corresponds to the mass shedding
(Keplerian) limit.  The value of $R_{\rm max}$
results from the condition that the frequency of a test
particle at circular equatorial orbit of radius $R_{\rm max}$
just above the equator of the {\it actual rotating star}  is
equal to 1122 Hz. It is interesting that the relation between $M$
and $R_{\rm eq}$ at this "mass shedding point" is extremely
well approximated by the formula for the orbital frequency for
a test particle  orbiting at  $r=R_{\rm eq}$ in the
Schwarzschild space-time created by a {\it spherical mass} $M$
(which can be replaced by a point mass $M$ at $r=0$). We denote
the orbital frequency of such a test particle by
$f^{\rm Schw.}_{\rm orb}(M,R_{\rm eq})$.
The formula giving the locus of points satisfying
$f^{\rm Schw.}_{\rm orb}(M,R_{\rm eq})=1122\;$Hz, represented by a dash
line in Fig.\ \ref{fig:MR-NS}, is
\begin{equation}
{1\over 2\pi}\left( {GM\over{ R_{\rm eq}}^3}\right)^{1/2}=1122~{\rm
Hz}~.
\label{eq:f-orb.1122Hz}
\end{equation}
This formula for the Schwarzschild metric coincides with that
obtained in Newtonian gravity for a point mass $M$. It
passes through (or extremely close to)  the open circles
denoting the actual mass shedding (Keplerian) configurations.
This  is  quite surprising
in view of rapid rotation and strong flattening of neutron star
at the mass-shedding point (see Fig.\ \ref{fig:shapesNS}).
Equation (\ref{eq:f-orb.1122Hz}) implies
\begin{equation}
R_{\rm max}=15.52\;\left({M\over 1.4\;M_\odot}\right)^{1/3}\;{\rm km}~.
\label{eq:Rmax.1122Hz}
\end{equation}

Out of ten EOSs of neutron star matter, two were chosen to represent
a soft (BPAL12) and stiff (GN3) extremes of the set. These two limiting
EOSs should  not  be considered as ``realistic'', but they are used just to
``bound'' the neutron star models from the soft and the stiff side.
 The curves representing allowed configurations rotating at
1122 Hz are  plotted in the $M-R_{\rm eq}$ plane
 in Fig.\ \ref{fig:MR-NS}.

The EOSs based on most realistic models
involving only nucleons (FPS, BBB2, DH, APR) result in monotonic
$M(R_{\rm eq})$ of very characteristic ``tilda-like'' shape.
For these EOSs we get $R_{\rm min}\approx 10\;$km and
$R_{\rm max}\approx 16\;$km.
 The value of $M(R_{\rm min})$ is significantly larger than $M(R_{\rm max})$,
 the difference ranging from $0.5\;M_\odot$ to $0.8\;M_\odot$.
The ratio of the kinetic to the gravitational energy $T/|W|$ is about
0.09-0.11 at the mass shedding limit at $R_{\rm max}$ and is
as low as 0.03-0.04 at $R_{\rm min}$.

 Four  EOSs are softened at high
 density either by the appearance of hyperons (GNH3, BGN1H1),
 or a phase transition (GMGSm, GMGSp). For them, the
 range of allowed masses is very narrow, $\sim 0.1\;M_\odot$,
 and moreover $M(R_{\rm min})<M(R_{\rm max})$. In spite of
   $(\partial M/\partial \rho_{\rm c})_{f=1122\;{\rm Hz}}<0$,
   configurations between $R_{\rm min}$ and $R_{\rm
 max}$ are stable with respect to small axi-symmetric
 perturbations, because they satisfy stability condition
 \begin{equation}
 \left({\partial M\over \partial \rho_{\rm c}}
 \right)_J>0~,
 \label{eq:ax-sym.stab}
 \end{equation}
where the derivative is calculated along a sequence with a
constant stellar angular momentum $J$
\citep{FriedmanISor1988}. Peculiar $M(R_{\rm eq})$ dependence
 is related to the fact that configurations belonging to
 the $R_{\rm min}<R_{\rm eq}<R_{\rm max}$  are {\it
supramassive}, i.e., their baryon mass $M_{\rm b}$ is larger
than the maximum allowable baryon mass for non-rotating stars,
$M_{\rm b}>M_{\rm b,max}^{\rm stat}$.
The range of allowed values of $R_{\rm eq}$ is not much
different from that for purely nucleon EOSs.
\begin{table}
\caption{Parameters of neutron stars rotating at $f=1122$ Hz
at the $R_{\rm min}$  termination (instability with respect to
axi-symmetric perturbations) and the $R_{\rm max}$ termination
(mass shedding instability)
of the $M-R_{\rm eq}$ line, for ten EOSs of
neutron star matter.}
\begin{center}
 \begin{tabular}[t]{c|c|c|c|c}
 \hline\hline
EOS & $M(R_{\rm min})$ & $R_{\rm min}$ &
$r_\mathrm{pole}/r_{\rm eq}$  & $T/|W|$\\
     & $(M_\odot)$ &  (km) &    & \\
     \hline
GN3$^a$& 2.309 & 12.77 & 0.813 & 0.057\\
APR$^b$& 2.274 & 10.47 & 0.896  & 0.038\\
DH$^c$& 2.138 & 10.50& 0.887  & 0.038 \\
GNH3$^a$& 2.119 & 12.67 & 0.804 & 0.055\\
BBB2$^d$& 1.996 & 9.928 & 0.896 & 0.034\\
FPS$^e$& 1.872 & 9.75 & 0.894 & 0.033\\
BG1H1$^f$& 1.699 & 9.99 & 0.876 & 0.032\\
BPAL12$^g$& 1.521 & 9.64 & 0.875 & 0.032\\
GMHSm$^h$& 1.481 & 9.65 & 0.874 & 0.029 \\
GMGSp$^i$& 1.473 & 9.60 & 0.875 & 0.028 \\
 \hline\hline
 EOS & $M(R_{\rm max})$ & $R_{\rm max}$ &
$r_\mathrm{pole}/r_{\rm eq}$  & $T/|W|$\\
     & $(M_\odot)$ &  (km) &    & \\
     \hline
GN3$^a$& 2.298 & 18.25 & 0.564 & 0.117\\
APR$^b$& 1.498 & 15.93 & 0.563 & 0.108\\
DH$^c$& 1.648 & 16.45 & 0.560 & 0.111 \\
GNH3$^a$& 2.163 & 17.54 & 0.583 & 0.103\\
BBB2$^d$& 1.473 & 15.85 & 0.563 & 0.107\\
FPS$^e$& 1.393 & 15.56 & 0.566 & 0.104\\
BG1H1$^f$& 1.818 & 16.92 & 0.568 & 0.103\\
BPAL12$^g$& 1.422 & 15.61 & 0.582 & 0.089\\
GMHSm$^h$& 1.594 & 16.19 & 0.582 & 0.086\\
GMGSp$^i$& 1.606 & 16.23 & 0.581 & 0.086\\
\hline
\hline
\end{tabular}
\end{center}
 {References for the EOSs: $^a$ - \citet{Glend1985};
 $^b$ - \citet{APR1998}; $^c$ - \citet{DH2001}; $^d$ - \citet{BBB1997},
for Paris nucleon-nucleon potential; $^e$ - \citet{PandRavFPS1989};
 $^f$ - \citet{BG1997}; $^g$ - \citet{Bombaci1995}; $^h$ - \citet{PonsKcond2000},
GM+GS model with $U_K^{\rm lin}=-130$~MeV and pure normal and kaon condensed
phases; $^i$ - \citet{PonsKcond2000}, GM+GS model with
 $U_K^{\rm lin}=-130$~MeV and mixed-phase state between the
 normal and kaon-condensed phases.}
\label{tab:RminRmax}
\end{table}
EOSs GMGSp and GMGSm describe nucleon matter with a first
order phase transition due to kaon condensation. The hadronic
Lagrangian is the same in both cases. However, to get GMGSp we
assumed that the phase transition takes place between two pure
phases and is accompanied by a density jump. On the contrary,
assuming that the transition occurs via a mixed state of two
phases, we get EOS GMGSm. This last situation prevails when the
surface tension between the two phases is below a certain
critical value. The actual value of the
surface tension is very poorly known, and therefore it is
comforting to find that both curves are nearly identical.

\section{Quark stars at 1122 Hz}
\label{sect:QS}
\begin{figure}[h]
\centering
\resizebox{3.5in}{!}{\includegraphics[]{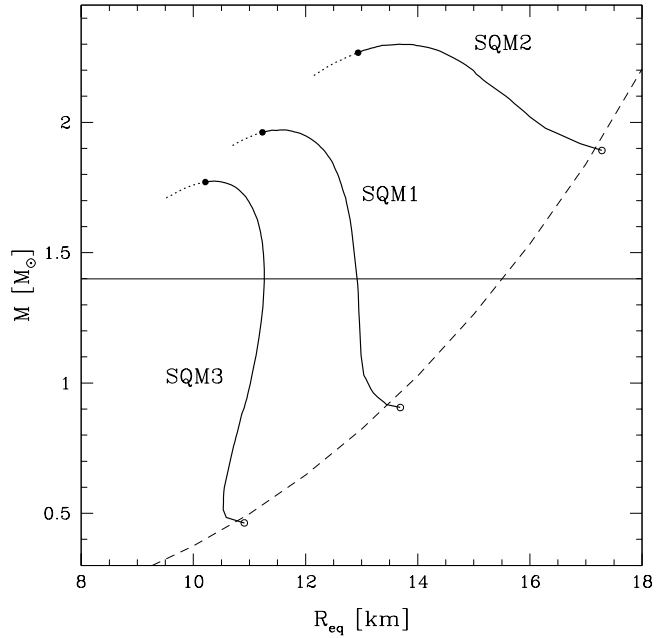}}
\caption{Gravitational mass, $M$, vs. circumferential equatorial radius,
$R_\mathrm{eq}$,    for strange stars with crust stably rotating at $f=1122$
Hz, for several EOS of strange matter. Notation as in Fig.\ \ref{fig:MR-NS}.
EOSs calculated using MIT bag model, with $\alpha_{\rm s}=0.2$ for all models
and $m_s c^2=200, 185, 205\;$MeV, $B=56, 45, 67\;{\rm MeV~fm^{-3}}$
for SQM1,2,3, respectively
(notation as in \citealt{HZS1986}).
 }
\label{fig:MR-SS}
\end{figure}
The possibility that the true ground state of hadronic matter
is  a self-bound plasma of u, d and s quarks was contemplated
since 1970s (\citealt{Bodmer1971}, \citealt{Witten1984}). If such a
state, called {\it strange matter}, is the true ground state
of hadronic matter, quark stars built entirely of strange
quark matter could exist (\citealt{Witten1984}, \citealt{HZS1986},
\citealt{AFO1986}). Could compact star in  XTE J1739-285 be
actually a quark star? An accreting quark star would have a crust
of normal matter and due to instabilities of thermonuclear burning
of accreted plasma it  could be an X-ray burster
(\citealt{HZQSacc1991}, \citealt{Page2005}). The mass of the crust on quark star is
at most $10^{-5}\;M_\odot$, because the density at the crust bottom
cannot exceed neutron drip density  \citep{AFO1986}.

As we see in Fig.\ \ref{fig:MR-SS}, the $M(R_{\rm eq})$ curves for quark
stars terminate at $R_{\rm max}$ which is rather well approximated by
Eq.\ \ref{eq:Rmax.1122Hz} (within 2\%).
The parameters of non-rotating strange stars are subject to the scaling laws with the value of the bag constant $B$
($M\propto B^{-1/2}$, $R\propto B^{-1/2}$) if the strange quark mass is scaled
 as $\propto B^{1/4}$ \citep{Zdunik2000}. The same scaling rules are valid also for
strange stars with crust provided that the mass of the crust is much smaller than
the mass of the strange core (see Eq.\ (7) in \citet{Zdunik2002} which results in
$R\propto B^{-1/2}$). For rotating stars the application of these scaling laws requires
also scaling of the rotational frequency with $f\propto B^{1/2}$ \citep{GourgSS1999}.
Thus these scaling laws cannot be directly applied to the situation presented if
Fig. \ref{fig:MR-SS} since the models SQM1, SQM2, SQM3 rotate with the same frequency
1122~Hz, and the condition $f_2=f_1\sqrt{B_2/B_1}$ is not fulfilled.

In contrast to hadronic EOSs, minimum mass allowed
for accreting quark star can be very low. However, if
we add the constraint of $M_{\rm max}^{\rm
stat}>2.1\pm 0.2\;M_\odot$ resulting from
the existence of a massive
slowly rotating pulsar \citep{Nice2005},
 we get $M(R_{\rm max})>1.4 \;M_\odot$. Let us notice that the
 values of $T/|W|$ at $R_{\rm max}$ are much larger than for
 neutron stars, 0.174 and  0.188 for SQM1 and SQM2, which makes
 these configurations secularly unstable with respect to the
 triaxial instability \citep{Gondek2003}, so that the actual
 value of $R_{\rm max}$ could be smaller and $M(R_{\rm max})$ could be
 larger.
\section{Discussion and conclusions}
\label{sect:discuss}
We constructed neutron star models rotating at 1122 Hz,
 suggested to be the rotation frequency of neutron star in
XTE J1739-285. At such frequency, rotation significantly affects neutron star
structure. The neutron star mass is larger then $1.4\;M_\odot$. The
equatorial (circumferential) radius is contained between
$R_{\rm min}=10-13\;$km and $R_{\rm
max}=15.52\;(M/1.4M_\odot)^{1/3}\;{\rm km}=16-18~{\rm km}$. The
formula for $R_{\rm max}(M)$ is very
precise, despite very strong deformation of neutron star at
the mass-shedding limit.

The mass of neutron star in XTE J1739-285 is between $M(R_{\rm max})$ and
$M(R_{\rm min})$. For EOSs assuming purely nucleon composition
of matter, $M(R_{\rm min})$ is
maximum allowable
mass at $f=1122\;$Hz
and $M(R_{\rm min})-M(R_{\rm max})\approx 0.5-0.7\;M_\odot$.

The situation is very different for EOSs with high-density
softening due to hyperons or due to a phase transitions. For
such EOSs $M(R_{\rm min})$ is slightly {\it smaller} than $M(R_{\rm
max})$, and the range of allowed masses is very narrow,
$|M(R_{\rm min})-M(R_{\rm max})|\approx 0.1\;M_\odot$.
Moreover, rotating neutron stars  are
then supramassive, with baryon
mass greater than the maximum allowable baryon mass for
nonrotating stars, $M_{\rm b}>M_{\rm b,max}^{\rm stat}$.
Their existence requires rapid rotation
and they collapse into black holes after spinning down.
However, because of a very narrow range of allowed masses,
possible presence of such a star in XTE J1739-285 would be a
result of ``fine-tuning'' in accretion process. This
could explain the rarity of submillisecond pulsars.
However, by turning the argument around, this implies that we are
very lucky to detect an extremely rare system! It should be
stressed that a measurement of  the mass $2.1\pm 0.2\;M_\odot$
of a {\it slowly
rotating} neutron star \citep{Nice2005} would anyway rule out
soft-core neutron stars not only in XTE J1739-285, but
generally.

Concerning neutron stars with a phase transition, as models of
XTE J1739-285, we found that their $M(R_{\rm eq})$ curves are
  nearly independent on whether phase
transition occurs between pure phases, or via a mixed-phase
state.

Could compact star in XTE J1739-285 be actually a strange star?
If so, the strange star should have an accreted crust of normal matter.
It remains to be studied, whether the observed X-ray bursts characteristics
could then be reproduced with such an exotic burster model.
In contrast to neutron stars, rotation at 1122 Hz alone would not
require a high $M$. However, if we additionally
require $M_{\rm max}^{\rm stat}>2.1\pm 0.2\;M_\odot$, then the range
of masses is narrower, $1.4-2.3\;M_\odot$.

 Constraints on the dense matter EOS, derived in the present
Letter,  are  {\it necessary} conditions resulting from the
existence of neutron star rotating at 1122 Hz, but they are
not {\it sufficient}.  Actually, neutron star in XTE J1739-285
acquired its rapid rotation  by accretion. Spin-up by
accretion to 1122 Hz leads to a stronger constraint on the EOS
than just stable rotation at this frequency. This problem was
studied for the 641 Hz pulsar by \cite{CSTacc1994}. Accretion
associated with spin can be followed as a line in the mass -
equatorial radius plane, and 1122 Hz should be reached before
the star becomes unstable either with respect to the mass
shedding or the  axi-symmetric instability. As an example, we
studied the accretion-driven spin up for the DH EOS. We found
that to spin up the star by accretion from a Keplerian disk to
$f=1122~$Hz, one needs to accrete  $0.37~M_\odot$ for an
initial mass $M_{\rm i}=1.4~M_\odot$.
For $M_{\rm i}=1.93~M_\odot$,  accretion of $0.25~M_\odot$ is
needed. For higher masses of initially
nonrotating star,  $1.93~M_\odot < M_{\rm i} < M_{\rm
max}=2.05~M_\odot$,  accretion leads to the stellar collapse into
a black hole before reaching rotation frequency of
1122~Hz. Our complete results on the
constraints resulting from the spin-up by accretion
will be presented in a separate paper.

When this Letter was being  completed, the preprint of
Lavagetto et al. (2006) appeared on the {\tt astro-ph}
server, in which neutron star rotating at 1122Hz is
used to derive one of several constraints on the dense matter EOS.
In contrast to the present Letter, Lavagetto et al. (2006) use
approximate ``empirical formula'' for the mass-shedding frequency
and get constraints in the mass-radius plane of non-rotating neutron stars
and strange stars.
\acknowledgements{ We thank an anonymous referee for raising
the point of spin-up by accretion.  This work was partially
supported by the Polish MNiI grant no. 1P03D.008.27. MB was
also partially supported by the Marie Curie Intra-european
Fellowship MEIF-CT-2005-023644 within the 6th European
Community Framework Programme.}



\vspace{-0.6cm}
\begin{thebibliography}{}
%
%
\bibitem[Akmal et al.(1998)]{APR1998}
Akmal A., Pandharipande V.R., Ravenhall D.G., 1998, Phys.Rev.
C, 58, 1804
%
\bibitem[Alcock  et al.(1986)]{AFO1986}
Alcock C., Farhi E., Olinto A.V., 1986,
ApJ, 310, 261
%
\bibitem[Backer et al.(1982)]{Backer1982}
Backer D.C., Kulkarni S.R., Heiles C., et al., 1982,
Nature, 300, 615
%
\bibitem[Balberg \& Gal(1997)]{BG1997}
Balberg S., Gal A., 1997, Nucl. Phys. A., 625, 435
%
\bibitem[Baldo et al.(1997)]{BBB1997}
Baldo M., Bombaci I., Burgio G.F., 1997, A
\& A, 328, 274
%
\bibitem[Bodmer(1971)]{Bodmer1971}
Bodmer A.R., 1971, Phys. Rev. D, 4, 1601
%
\bibitem[Bombaci(1995)]{Bombaci1995}
Bombaci I., 1995, in: Perspectives on Theoretical Nuclear Physics,
ed. by I. Bombaci, A. Bonaccorso, A. Fabrocini, et al. (Pisa: Edizioni ETS),
 p. 223
%
\bibitem[Bonazzola et al.(1993)]{BGSM1993}
Bonazzola S., Gourgoulhon E., Salgado M., Marck J.-A., 1993, A
\& A, 278, 421
%
\bibitem[Cook et al.(1994a)]{CST1994}
Cook G. B., Shapiro S.L., Teukolsky S.A. 1994a ApJ, 424, 823
%
\bibitem[Cook et al.(1994b)]{CSTacc1994}
Cook G. B., Shapiro S.L., Teukolsky S.A. 1994b ApJ, 423, L117
%
\bibitem[Douchin \& Haensel(2001)]{DH2001}
Douchin F., Haensel P. 2001, A \& A, 380, 151
%
\bibitem[Friedman et al.(1988)]{FriedmanISor1988}
Friedman, J.L., Ipser, J.R., Sorkin, R.D., 1988,
ApJ, 325, 722
%
\bibitem[Glendenning(1985)]{Glend1985}
Glendenning, N. K. 1985, ApJ, 293, 470
%
%
\bibitem[Gondek-Rosi{\'n}ska et al.(2003)]{Gondek2003}
Gondek-Rosi{\'n}ska D., E. Gourgoulhon E., Haensel P., 2003,
A \& A, 412, 777
%
\bibitem[Gourgoulhon et al.(1999)]{GourgSS1999}
Gourgoulhon, E., Haensel, P., Livine, R., Paluch, E.,
Bonazola, S., Marck, J.-A., 1999,
A\&A, 349, 851
%
\bibitem[Haensel et al.(1986)]{HZS1986}
Haensel P., Zdunik J. L., Schaeffer R., 1986, A\&A, 160, 251
%
\bibitem[Haensel \& Zdunik(1991)]{HZQSacc1991}
Haensel P., Zdunik J. L., 1991, Nucl. Phys. B (Proc. Suppl.), 24 B, 139
%
\bibitem[Haensel et al.(1995)]{HSB1995}
Haensel P., Salgado M., Bonazzola S., 1995, A\&A, 296, 745
%
\bibitem[Hessels  et al.(2006)]{Hessels2006}
Hessels J.W.T., Ransom S.M., Stairs I.H., et al., 2006,
Science, 311, 1901
%
%
\bibitem[Kaaret et al.(2006)]{Kaaret2006}
Kaaret P., Prieskorn Z., in't Zand J.J.M., Brandt S., Lund N.,
Mereghetti S., Goetz D., Kuulkers E., Tomsick J.A. 2006.,
 ApJ Letters, submitted, {\tt astro-ph/0611716}
%
%
\bibitem[Lattimer \& Prakash(2004)]{LP2004}
Lattimer J.M., Prakash M. 2004, Science, 304, 536
%
%
\bibitem[Lavagetto et al.(2006)]{Lavagetto2006}
Lavagetto G., Bombaci I., D'Ai A., Vidana I., Robba
N.R., 2006,  ApJ Letters, submitted,
{\tt astro-ph/0612061}
%
%
\bibitem[Nice et al.(2005)]{Nice2005}
Nice D., Splaver E.M., Stairs I.H., et al., 2005,
 ApJ, 634, 1242
%
%
\bibitem[Nozawa et al.(1998)]{Nozawa1998}
Nozawa T., Stergioulas N., Gourgoulhon E., \& Eriguchi Y.,
1998, A\&A, 139, 431
%
%
\bibitem[Page \& Cumming(2005)]{Page2005}
Page D., Cumming A., 2005, ApJ, 635, L157
%
\bibitem[Pandharipande \& Ravenhall(1989)]{PandRavFPS1989}
Pandharipande V.R., Ravenhall D.G. 1989, in Proc. NATO
Advanced Research Workshop on nuclear matter and heavy ion
collisions, Les Houches, 1989, ed. M. Soyeur et al. (Plenum,
New York, 1989), 103
%
\bibitem[Pons et al.(2000)]{PonsKcond2000}
Pons J.A., Reddy S., Ellis P.J., Prakash M., Lattimer J.M.,
2000, Phys. Rev. C, 62, 035803
%
\bibitem[Salgado et al.(1994)]{Salgado1994}
Salgado M., Bonazzola S., Gourgoulhon E., Haensel P.,
1994, A \& A 108, 455
%
\bibitem[Shapiro \& Teukolsky(1983)]{ShapTeuk1983}
Shapiro S.L., Teukolsky S.A. 1983, Black Holes, White Dwarfs,
and Neutron Stars (Wiley, New York)
%
\bibitem[Shapiro et al.(1983)]{STW1983}
Shapiro S.L., Teukolsky S.A., Wasserman I., 1983, ApJ, 272,
702
%
\bibitem[Witten(1984)]{Witten1984}
Witten, E.,  1984,
Phys. Rev. D, 30, 272
%
\bibitem[Zdunik(2000)]{Zdunik2000}
Zdunik J.L., 2000, A\&A, 359, 311
%
\bibitem[Zdunik(2002)]{Zdunik2002}
Zdunik J.L., 2002, A\&A, 394, 641

\end{thebibliography}
\end{document}